\documentclass[a4paper, 12pt]{article} % Font size (can be 10pt, 11pt or 12pt) and paper size (remove a4paper for US letter paper)
\usepackage[right=1.1in,left=1.1in,top=1.1in,bottom=1.1in]{geometry}
\usepackage{graphicx} 
\usepackage{braket}
\usepackage{mathrsfs}
\usepackage[T1]{fontenc} % Required for accented characters
\usepackage{amsmath}
\usepackage{pxfonts}
\usepackage[T1]{fontenc}
\usepackage{amssymb}
\usepackage{natbib}
\usepackage{epstopdf}
\usepackage{setspace}
\usepackage{enumitem}
\usepackage{sectsty}
\usepackage{wrapfig} % Allows in-line images
\sectionfont{\large}
\bibpunct{(}{)}{,}{a}{}{,}
\usepackage{tikz}   %TikZ is required for this to work.  Make sure this exists before the next line
\usepackage{tikz-3dplot} %requires 3dplot.sty to be in same directory, or in your LaTeX installation
\newcommand{\qed}{\nobreak \ifvmode \relax \else
      \ifdim\lastskip<1.5em \hskip-\lastskip
      \hskip1.5em plus0em minus0.5em \fi \nobreak
      \vrule height0.75em width0.5em depth0.25em\fi}

\usepackage{bbm}
\usepackage{MnSymbol}
\usepackage[all]{xy}

\usepackage{xcolor,colortbl,array,amssymb}

\usepackage{multirow}

\makeatletter
\title{Bell's Theorem, Quantum Probabilities, and Superdeterminism} % Subtitle

\author{Eddy Keming Chen\thanks{Department of Philosophy,  University of California, San Diego, 9500 Gilman Dr, La Jolla, CA 92093-0119. Website: www.eddykemingchen.net. Email: eddykemingchen@ucsd.edu  }}

\date{\today \vspace{10pt} \\ Forthcoming in Eleanor Knox and Alastair Wilson (eds.), \\  \emph{The Routledge Companion to the Philosophy of Physics} } % Date

\begin{document}
\bibliographystyle{plain}

\maketitle % Print the title section

%----------------------------------------------------------------------------------------
%	ABSTRACT AND KEYWORDS
%----------------------------------------------------------------------------------------

%\renewcommand{\abstractname}{Summary} % Uncomment to change the name of the abstract to something else
%\textbf{ Thanks for your interests. This is a very rough work-in-progress version. Any comments are welcome. Don't worry too much about the technical details. \S 2 and \S 3 are mostly general philosophy of science. \S 3 contains some technical stuffs. If you do care about them, I'd be grateful for your feedback.}

\begin{abstract}
In this short survey article, I discuss Bell's theorem and some strategies that attempt to avoid the conclusion of non-locality. I  focus on two that intersect  with the philosophy of probability: (1) quantum probabilities and (2) super-determinism. The issues they raised not only apply to a wide class of no-go theorems about quantum mechanics but are also of general philosophical interest. 

\end{abstract}

\hspace*{3,6mm}\textit{Keywords: Bell's theorem, non-locality, quantum probabilities, Kolmogorov  axioms, super-determinism, simplicity, complexity, initial conditions of the universe}   % Keywords

\begingroup
\singlespacing
\tableofcontents
\endgroup

\vspace{30pt} % Some vertical space between the abstract and first section

%----------------------------------------------------------------------------------------
%	ESSAY BODY
%----------------------------------------------------------------------------------------

%Blah blah \citet{PaulNREWM} \citep{PaulNREWM}  

%------------------------------------------------

%\section*{Summary}

%\begin{enumerate}[label= (\arabic*)]
%\item $4a+5d=6c$
%\item $2^{a+b}$
%\end{enumerate}

 %$$3c+\sqrt[3]{d}$$ 

%------------------------------------------------
\nocite{sep-probability-interpret}

\section{Introduction}

As early as the beginning of quantum mechanics, there have been numerous attempts to prove impossibility results or ``no-go'' theorems about quantum mechanics. They aim to show that certain plausible assumptions about the world are impossible to maintain given the predictions of quantum mechanics, which can and have been empirically confirmed. Some of them are more significant than others. Arguably, the most significant  is J. S. Bell's (\citeyear{bell1964einstein}) celebrated  theorem of non-locality: given plausible assumptions, Bell shows that, in our world, events that are arbitrarily far apart can instantaneously influence each other.

Bell's theorem is most significant because its conclusion is so striking and its assumptions so innocuous that it  requires us to radically change how we think about the world (and not just about quantum theory).   

Before Bell's theorem, the picture we have about the world is like this: physical things interact only locally in space. For example, a bomb dropped on the surface of Mars will produce immediate physical effects (chemical reactions, turbulences, and radiations) in the immediate surroundings;  the event will have (much milder) physical effects on Earth only at a later time, via certain intermediate transmission between Mars and the Earth. More generally, we expect the world to work in  a local way such that events arbitrarily far apart in space cannot instantaneously influence one another. This picture is baked into classical  theories of physics such as Maxwellian electrodynamics and (apparently) in relativistic spacetime theories. 

 After Bell's theorem, that picture is untenable. Bell proves that Nature is non-local if certain predictions of quantum mechanics are correct. Many experimental tests (starting with \cite{aspect1982experimentalA} and \cite{aspect1982experimentalB}) have been performed. They confirm over and over again the predictions of quantum mechanics. Hence, we should have extremely high confidence in the conclusion that Nature is non-local: events that are arbitrarily far apart in space can instantaneously influence each other.  (In the relativistic setting, it amounts to the conclusion that events that are space-like separated can influence each other.) 

However, not everyone is convinced. In fact, there are still disagreements about what Bell proved and how general the result is. Some disagreements can be traced to misunderstandings about the assumptions in the proof. Others may be due to more general issues about scientific explanations and the standards of theory choice. 

There are many good articles and books about Bell's theorem. (For example, see \cite{maudlin2011quantum, maudlin2014bell}, \cite{Goldstein2011Bell}, and \cite{sep-bell-theorem}.) In this short article, I would like to focus on two strategies that attempt to avoid the conclusion of non-locality. They are about (1) quantum probabilities and (2) super-determinism, both having to do, in some ways, with the philosophy of probability. First, I argue that solving the problem by changing the axioms of classical probability theory is a non-starter, as Bell's theorem only uses frequencies and  proportions that obey the rules of arithmetic. Moreover, this point is independent of any interpretation of probability (such as frequentism). Second, I argue that a super-deterministic theory may end up requiring an extremely complex initial condition, one that deserves a much lower prior probability than its non-local competitors.   Since both issues can be appreciated without much  technical background and have implications for other subfields of philosophy, I will try to present them in a non-technical way that is accessible to non-specialists.   The lessons we learn from them also apply to the more recently proven theorem (2012) of  Pusey, Barrett, and Rudolph about the reality of the quantum state, which is in the same spirit as Bell's theorem. (Their theorem says that, under plausible assumptions,  quantum states represent states of reality rather than merely certain knowledge about reality.)

%Here is the plan. In \S2, I  discuss a version of Bell's theorem. In \S3, I discuss a response to avoid non-locality by changing the axioms of classical probability theory. I argue that it is unlikely to work. In \S4, I discuss another response---known as \emph{super-determinism}---that rejects a statistical independence assumption. I argue that it probably leads to a highly complex theory that deserves only a credence that is much lower than that of existing non-local theories.

\section{Bell's Theorem}

There are many versions of Bell's theorem and Bell inequalities. For illustration, in this section, we discuss a version of them by adapting a simple example involving perfect correlations discussed in \cite{maudlin2011quantum}\S1. (Another simple example, involving perfect anti-correlations, can be found in \cite{albert1994quantum}\S3.)

Under certain physical conditions, the calcium atom can emit a pair of photons that travel in opposite directions: left and right. We have labs that can realize such conditions. In  this situation, we can set up polarizers on the left and on the right, as well as devices on both sides that detect photons that happen to pass through the polarizers.  If a photon is absorbed by a polarizer, then the photon detector placed behind the polarizer will detect nothing. (Here we assume that the photon detectors are 100\% reliable. The idealization can be relaxed, and analyses have been done to show that  the differences do not change the conclusion we want to draw.) Further, we can arrange the polarizers to be pointing in any direction on a particular plane.  Each direction is representable by a number between 0 and 180, corresponding to the clockwise angle of the polarizer away from the vertical direction. Since either polarizer receives exactly one incoming photon, we say that the pair of photons agree if they either both passed or both got absorbed by the polarizers (so the photon detectors on both sides clicked or neither did); they disagree if one passed but the other got absorbed (so exactly one photon detector clicked). 

When we carry out the experiments, say, by using 100,000 pairs of photons, quantum mechanics predict that we would observe the following:  

\begin{itemize}
	\item Prediction 1: If the left polarizer and the right polarizer point in the same direction, 100\% of the pairs  agree. 
	\item Prediction 2: If the left polarizer and the right polarizer differ in direction by 30 degrees,   25\% of the pairs  disagree. 
	\item Prediction 3: If the left polarizer and the right polarizer differ in direction by 60 degrees,  75\% of the pairs disagree. 
\end{itemize}
(The situation is  a bit simplified. In actual experiments, the empirical frequencies  will be approximately 25\% and approximately 75\% respectively and will increasingly approach them  as we carry out more trials.) 
In the end, these statistics  will be shown to clash with a plausible hypothesis of locality: 

\begin{description}
	\item[Locality] Events  arbitrarily far away cannot instantaneously influence each other. 
\end{description}

Bell shows that the conjunction of Locality and the predictions of quantum mechanics leads to a contradiction. There are two parts in Bell's argument.  The first part  is based on the argument of \cite{einstein1935can}, also known as the EPR argument. 

In the context of our example, the EPR argument can be summarized as follows. First, the photon traveling to the left and the photon traveling to the right can be separated arbitrarily far away. Second, we can always place a polarizer in the path of  the photon on the left and another in the path of the photon  on the right. Third, according to Prediction 1, if the two polarizers point in the same direction, the pair of photons always agree, however far away they are from each other. Moreover, if we first measure the photon on the left and find that it passed the polarizer on the left, then we do not even  need to measure the photon on the right if the polarizer on the right points in the same direction; we know the result---it will pass the polarizer on the right. Assume Locality: what happens to the photon on the (distant) left cannot instantaneously influence  the photon on the (distant) right. So there is already a fact of the matter, before measurement, about the result on the right. Hence, Locality implies that there are facts of the matter about the polarization direction of  the photon on the left and the photon on the right. In other words, their values of polarizations are predetermined. 

Here is another way to see this. Given Prediction 1, since there is no way to ``know'' the directions of the two polarizers, the photons must already agree, even inside the calcium atom, how they would react to the polarizers come what may. That is, they must already agree whether to both pass or both get absorbed for polarizers pointing to any particular angle. For example, they must ``agree'' how to react when facing polarizers pointing at 0 degrees, when facing  polarizers pointing at 30 degrees, when facing polarizers pointing at 60 degrees, and so on. Otherwise they would not be able to satisfy Prediction 1. However, such predetermined facts are not included in the quantum mechanical description using a wave function. So somehow these facts will be encoded in further parameters going beyond quantum theory. Indeed, the EPR argument aims to show that Locality implies that quantum mechanics is an incomplete description of Nature. (A famous example of a theory that adds additional parameters is the de Broglie-Bohm theory, but it is manifestly non-local in the particle dynamics. So it is not an example of the kind of local completion of quantum mechanics that EPR look for. Nevertheless, the non-local character of the de Broglie-Bohm theory was one of the motivations for Bell to investigate the generality of non-locality. See  \cite{bell1964einstein}\S1 and \cite{bell2001introduction}.)

In short, what was shown by EPR and used in Part I of Bell's argument is the following: 

\begin{itemize}
	\item[Part I] Locality \& Quantum Predictions $\Longrightarrow$ Predetermined Values 
\end{itemize}

In Part II, Bell shows the following:

\begin{itemize}
	\item[Part II] Predetermined Values  \& Quantum Predictions $\Longrightarrow$ Contradiction  
\end{itemize}

We will see that predetermined values and quantum predictions lead to a contradiction with the laws of arithmetic (regarding addition, multiplication, and fraction). Recall that there are  facts of the matter about the polarization properties of the pair of photons. But there are still two possibilities for each angle. For example, for polarizers pointing at 30 degrees, there can be two alternatives: both pass and both get absorbed. To simplify the example, we assume that the directions of the polarizers have only  three choices (say, limited by the turning knobs on the devices): 0 degrees, 30 degrees, and 60 degrees. Then for each choice of the angle of polarizer, there can be two possibilities for the pair: both pass (P) or both get absorbed (A). For example, they may both instantiate $P_{30}$, which means they will both pass if the polarizer is pointing at a 30 degrees angle; they may both instantiate $A_{60}$, which means they will both get absorbed if the polarizer is pointing at a 60 degrees angle.  Since $2^3 = 8$, there are exactly eight choices for the assignments of properties in the two photons.

\centerline{
\begin{tabular}{ | l | c | c  | c | c|}
  \hline
  \multicolumn{5}{|c|}{Eight Possible Assignments of Properties} \\
  \hline
   & Left Photon & Right Photon & Feature  & Percentage  \\
\hline
  (1) & $P_{0} , P_{30} , P_{60}$   & $P_{0} , P_{30} , P_{60}$  & \multirow{2}*{X} &\multirow{2}*{$\alpha$\%} \\
  (2) & $A_{0} , A_{30} , A_{60}$   & $A_{0} , A_{30} , A_{60}$  &  &  \\ \hline
  (3) & $A_{0} , P_{30} , P_{60}$   & $A_{0} , P_{30} , P_{60}$  & \multirow{2}*{Y} &   \multirow{2}*{$\beta$\%}  \\
  (4) &  $P_{0} , A_{30} , A_{60}$   & $P_{0} , A_{30} , A_{60}$  &  & \\ \hline
  (5) &  $P_{0} , A_{30} , P_{60}$   & $P_{0} , A_{30} , P_{60}$  & \multirow{2}*{Z}  &   \multirow{2}*{$\gamma$\%} \\
  (6) & $A_{0} , P_{30} , A_{60}$   & $A_{0} , P_{30} , A_{60}$  &  &  \\ \hline
  (7) & $P_{0} , P_{30} , A_{60}$   & $P_{0} , P_{30} , A_{60}$  & \multirow{2}*{W}  &   \multirow{2}*{$\delta$\%} \\
  (8) & $A_{0} , A_{30} , P_{60}$   & $A_{0} , A_{30} , P_{60}$   &  &    \\
    \hline
\end{tabular}
   }
To satisfy Prediction 1, different pairs of photons can choose exactly one of these eight assignments. If a pair does not choose among these eight, then it can violate experimental results. 

The eight assignments can be put in four groups as indicated in the table. Let us label the four groups with features X, Y, Z, and W, which we mention again in \S3.  Now suppose we have  a large number of pairs of photons emitted from a collection of calcium atoms. (The larger the number, the closer empirical frequencies will approach the predicted percentages.) Assuming Locality, each pair must adopt one of the eight assignments listed above. Let $\alpha$ be the percentage of pairs that realizes either (1) or (2); $\beta$ be the percentage of pairs that realizes either (3) or (4); $\gamma$ be the percentage of pairs that realizes either (5) or (6); and $\delta$ be the percentage of pairs that realizes either (7) or (8). By the laws of arithmetic, 
%\begin{itemize} \item Feature X:  the pair agrees no matter what.	\item Feature Y:  the pair disagree when exactly one of the two photons meets a polarizer placed at 0 degrees. 	\item Feature Z: the pair disagree when exactly one of the two photons meets a polarizer placed at 30 degrees. 	\item Feature W: the pair disagree when exactly one of the two photons meets a polarizer placed at 60 degrees. \end{itemize}

% and satisfy one of the four features X--W. Let $\alpha$ be the number of pairs that satisfy Feature X; $\beta$ be the number of pairs that satisfy Feature Y; $\gamma$ be the number of pairs that satisfy Feature Z; and $\delta$ be the number of pairs that satisfy Feature W. By the laws of arithmetic, 

\begin{equation}
	\alpha+\beta+\gamma+\delta=100
\end{equation}
Moreover, each percentage number must be non-negative. In particular,

\begin{equation}
	\gamma \geq 0
\end{equation}
Therefore, 

\begin{equation}
	\gamma + \delta + \beta + \gamma \geq \beta + \delta
\end{equation}

Unfortunately, this is inconsistent with the conjunction of Prediction 2 and Prediction 3. According to Prediction 2, if the the angles of the polarizers on the two sides differ by 30 degrees, then we find photon disagreement  25\% of the time.   We run the large number of pairs of photons with  the left polarizer pointing to 0 and the right pointing to 30. By inspection of the table, we know that pairs realizing assignments (1) and (2) will agree. So we know that  $\alpha$ percent of the pairs agree. Moreover, we know that pairs realizing assignments (7) and (8) will also agree. That is another $\delta$ percent of pairs that agree. The only pairs that disagree will be those realizing assignments (3), (4), (5), and (6). That is $\beta+\gamma$ percent  pairs that disagree. Hence,

\begin{equation}
	\beta+\gamma = 25
\end{equation}

Similar considerations apply when we set the left polarizer at 30 degrees and the right at 60 degrees. Then,

\begin{equation}
	\gamma + \delta= 25
\end{equation}

According to Prediction 3, if the the angles of the left and the right polarizers differ by 60 degrees, in our example that is when one is pointing at 0 and the other 60, then pairs of photons disagree 75\% of the time. All disagreements come from photon pairs that realize  assignments (3), (4), (7), and (8). Hence, 

\begin{equation}
	\beta+\delta = 75
\end{equation}

From the above three equations, since 50 is smaller than 75, we can conclude that

\begin{equation}
	\gamma + \delta + \beta + \gamma < \beta + \delta. 
\end{equation}
But equation (7) is inconsistent with equation (3). We have arrived at a contradiction. Hence, the second part of Bell's argument is established. Together, Part I and Part II imply: 
\begin{itemize}
	\item[ ] Locality  \& Quantum Predictions $\Longrightarrow$ Contradiction  
\end{itemize}
Since quantum predictions have been confirmed to an extremely high degree, we should have very high confidence that Locality is refuted and that Nature is non-local. (Here we take quantum predictions to be statistical---regarding empirical frequencies---rather than probabilistic.) Of course, we have made some implicit assumptions in the derivation:
\begin{itemize}
	\item[(A)] The rules of inferences obey classical logic.
	\item[(B)] The laws of arithmetic are true.
	\item[(C)] Frequencies and  proportions obey the laws of arithmetic. 
	\item[(D)] There are no conspiracies in nature. 
\end{itemize}
Strictly speaking, it is only by assuming (A)---(D) can we derive the contradiction from Locality and Quantum Predictions. We will return to these implicit assumptions in the next two sections. (Another assumption is the idea that each experimental outcome is unique and definite, which is denied in the Many-Worlds interpretation. See \S7 for further readings. One can label this as the fifth assumption. However, this assumption is arguably already contained in our description of Quantum Predictions about empirical frequencies. If experimental outcomes are not definite,  empirical frequencies wouldn't even make sense unless we state them in a different way, such as by pairing certain outcomes into a single branch and using ``branch-weighted'' frequencies.)

In this section, we have presented one version of Bell inequalities (in equation (3)) and explained how it is violated by the predictions of quantum mechanics (in equation (7)). (Bell's own version (1964) uses perfect anti-correlation and is stated in terms of expectation values. \cite{clauser1969proposed} provides a generalization of Bell's result that allows imperfect correlations.)

\section{Quantum Probabilities to the Rescue?}

Perhaps due to the significance of Bell's theorem, there have been many attempts that try to avoid the conclusion of non-locality  by identifying some other ``weak link'' in the argument. (For some examples, see Further Readings in \S7.) That is surprising,  since the other assumptions are quite innocuous and \emph{a priori}, as illustrated by the previous example. 

One purported ``weak link'' is associated with the ``implicit assumptions'' about classical probability theory. One might suspect that   the derivations of  Bell's theorem  require substantive assumptions about the nature of probability.  Probability is notoriously difficult  to understand. Hence, there may be room to revise our classical theory of probability  given empirical data.  The suggestion is that, instead of rejecting Locality, we can modify (or generalize) the classical axioms and algebraic structure of Kolmogorov probability theory to avoid the contradiction. (For example, see  \cite{fine1982hidden},  \cite{fine1982joint} and \cite{pitowsky1989quantum}.) 

However, the previous example serves as a counterexample. In the argument of \S2,  assumptions of classical probability theory do not even occur. Nor do they implicitly play any essential role. All we ever needed were  proportions and how they arithmetically interact with each other (addition, multiplication, subtraction, and division). For example, Predictions 1, 2, and 3 are formulated in terms of  percentages of pairs of photons. The four groups of  possible assignment of properties receive percentages $\alpha, \beta, \gamma,$ and $\delta$. We call them ``percentages,'' which may remind readers of probabilities. But in our argument they merely represent  proportions. To say that $\alpha$ percent of the pairs realize property assignments (1) or (2) is to say that the number of pairs having those properties is exactly $\alpha$ per 100 pairs. If we have 100,000 pairs in total in the collection, then that amounts to $1000 \times \alpha$ pairs. 

Since the percentages $\alpha, \beta, \gamma,$ and $\delta$ represent  proportions,  it is in their nature that they obey the laws of arithmetic, and their bearers (property assignments (1)-(8)) obey the rules of Boolean algebra. (\cite{tumulka2015assumptions} makes a similar point.)  The fact that we are assuming, in the conditional proof, they have hidden properties does not matter at all. As such,   proportions   obey the axioms governing how we should count a finite number of things, which obey the Kolmogorovian axioms, which may also govern probabilities (according to some interpretations of probability). Nevertheless, that does not make  proportions subject to various interpretational issues as probability does.  Many other concepts also satisfy Kolmogorovian axioms, including as  mass, length, and volume of finite physical objects. Neither are they subject to the interpretational controversies surrounding the concept of probabilities. Probability faces a wide range of interpretational puzzles, and it is controversial what its axioms ought to be. Still, there are no similar difficulties with concepts of mass, length,  volume,  frequencies, or proportions.

Why is it in the nature of frequencies and  proportions to obey the laws of arithmetic or  counting finite number of things?   This may seem like a question in the philosophy of mathematics. Fortunately, we do not need to settle those controversies to answer that question for our purposes here.  The discussion about non-classical probability spaces and  Bell's theorem is sometimes highly technical, and different proposals  have been suggested to understand violations of the rules of  Boolean algebra and Kolmogorov axioms. For our purposes we can distill the central intuitions using the concrete example of \S2. Suppose we have a large collection of photon pairs adequately prepared. Consider four features that each photon pair can have---X, Y, Z, and W---that are mutually exclusive and jointly exhaustive, and consider the following propositions:

\begin{itemize}
	\item[(i)] The percentage of photon pairs having exactly one of the four features is $100$\%. 
	\item[(ii)] The percentage of photon pairs having feature Z is non-negative. 
	\item[(iii)] The percentage of photon pairs having either Y or W is the sum of the percentage of photon pairs having Y and the percentage of photon pairs having W. 
	\item[(iv)] The sum of percentage of photon pairs having the property (Y or Z) and the percentage of photon pairs having the property (Z or W) is well defined---a non-negative number. 
\end{itemize}
Can these propositions be false? In particular, can they fail in the following ways?
\begin{itemize}
	\item[(i')] The percentage of photon pairs having exactly one of the four features is $115$\%. 
	\item[(ii')] The percentage of photon pairs having feature Z is $-5$\%. 
	\item[(iii')] The percentage of photon pairs having either Y or W is less than the sum of the percentage of photon pairs having Y and the percentage of photon pairs having W. 
	\item [(iv')]The sum of percentage of photon pairs having the property (Y or Z) and the percentage of photon pairs having the property (Z or W) does not exist. 
\end{itemize}
It is  \emph{a priori} that   propositions (i)---(iv) cannot be false while  propositions (i')---(iv') cannot be true. Propositions such as (i)---(iv) are sufficient to prove the violation of a Bell inequality (equation (3)) in \S2.  They are not dependent on any substantive theory or axioms about probabilities, because they are about  proportions and not  about probabilities. We do not need to appeal to assumptions about the nature of probabilities to prove that Nature is non-local.

A potential misunderstanding is that, to say the thing we just said, we must be endorsing a particular interpretation of probability---frequentism, according to which probabilities boil down to long-run frequencies. But that is a mistake. We can make judgments about those eight propositions without endorsing any particular interpretation of probability. To evaluate them, we do not have to settle the debate among subjectivism, frequentism, and the propensity interpretations. For example, one can be a subjective Bayesian about probabilities and still accept that   frequencies, percentages, and  proportions obey propositions (i)---(iv). One can even adopt the view that the actual axioms governing real probabilities are non-Kolmogorovian and involving non-Boolean algebra without denying that  frequencies and proportions obey the rules of arithmetic. (Moreover, the actual evidence we use to support quantum theory consists in empirical frequencies, which obviously obey the classical probability axioms.)

% Given the close connection between laws of arithmetic and rules of logic, it is hard to see how change in the former without change to the latter. So any attempt to use ``quantum'' probability axioms may also lead to the use of non-classical logic, an option that is much less desirable than just giving up classical probability theory. 

However, not everyone would agree with our assessment. \cite{fine1982hidden, fine1982joint}  and \cite{pitowsky1989quantum} seem to suggest it remains possible to save locality by revising classical probability theory. (See \cite{malament2006notes} for a  clear introduction to this project.  \cite{feintzeig2015hidden} demonstrates further mathematical constraints.)   The project has led to important and beautiful mathematical results that can shed light on the mathematical structures of impossibility theorems. Nevertheless, if the above analysis is correct,  then the project of avoiding non-locality by revising probability axioms is a non-starter; it cannot  get off the ground, no matter how ingenious or elegant the models of non-classical probability spaces are.  No matter what changes we make to classical probability theory, they do not affect the conclusion of non-locality. The argument for non-locality does not rely on classical probability theory. We only need to use rules for counting relative frequencies and proportions.

Quantum probability (as an alternative to classical probability) is  related to quantum logic (as an alternative to classical logic). Some people who want to keep classical logic may nonetheless be open to revise the axioms of probability to make room for locality. But as we just discussed, it is the axioms governing  frequencies and  proportions that need to be revised if one goes that route. Since  they obey  the axioms of arithmetics, and since the latter are closely related to logic, it is hard to see how to pursue this route without also revising logic in some way. (See \cite{sep-qt-quantlog} for a survey of quantum logic and quantum probability theory.)

Therefore,   we cannot save Locality by changing the axioms governing classical probability theory. Which probability theory is correct is an important question in the philosophy of probability but it is irrelevant to the question whether Nature is non-local.

\section{Escape with Super-Determinism?}

Another purported ``weak link'' in Bell's argument is associated with the assumption of statistical independence.  The strategy is to allow systematic violations of statistical independence in favor of ``super-deterministic'' theories. (This is sometimes labeled as ``conspiratorial theories.'')   In this section we will try to understand what the strategy is and what difficulties it faces. 

In \S2, we assumed that the direction of the polarizer can be set independently of the collection of incoming photon pairs. We can, for example, use a mechanical device that randomly selects (say, based on certain digits of $\pi$) among the three choices---pointing at 0 degrees, 30 degrees, and 60 degrees. That assumption---statistical independence---seems fundamental to scientific experimentation. Another way to see it is in terms of random sampling. Given any collection of photon pairs adequately prepared, and after the experimental set up is completed,  we can perform random sampling on the collection and obtain a sub-collection that  reflects the same statistical profile as the overall collection and any other sub-collection so randomly chosen. That is, if the sub-collection is such that 25\% of them would disagree when pairs of photons pass through polarization filters that differ by 30 degrees, then the whole collection (and other randomly chosen sub-collection) would also have that property. In other words, the choice of the sub-collections can be made statistically independent of the experimental setup. Statistical independence enables us to apply the conjunction of Prediction 1, Prediction 2, and Prediction 3 to the collection as a whole (and to each sub-collection) and to deduce equations (4), (5), and (6), from which we derive a contradiction with inequality (3).  

Without assuming statistical independence, the inference is not valid. We can construct an example in which the quantum predictions are all satisfied during experiments but there is no contradiction. Suppose we have  100,000 photon pairs to start with. Each photon pair realizes one of the eight assignments listed in the table. Suppose further that $\alpha=\beta=\gamma=\delta=25.$ We have three experimental setups:

\begin{itemize}
	\item[(A)] Left polarizer at 0 degrees, right polarizer at 30 degrees.
	\item[(B)] Left polarizer at 30 degrees, right polarizer at 60 degrees.
	\item[(C)] Left polarizer at 0 degrees, right polarizer at 60 degrees.
\end{itemize}
From  the collection of 100,000 photon pairs, we choose  three sub-collections---(a), (b), and (c)---each with exactly 100 photon pairs. It turns out that, when we send (a) through (A),  25\% of them disagree; when we send (b) through (B),   25\% of them disagree; when we send (c) through C,  75\% of them disagree. (As before, this is an idealization. The fractions get closer to these numbers when we run the trials with more pairs.) This can be realized in the following way. In (a), 25 pairs are of type (3) and the rest are of type (1); in (b), 25 pairs are of type (5), and the rest are of type (1); in (c), 75 pairs are of type (7) and the rest are of type (1).  That is, sub-collection (a) has exactly the kind of statistical profile required to be in agreement with quantum predictions for experiment (A);  sub-collection (b) for (B); and the sub-collection (c) for (C). Hence, each sub-collection has the ``right'' statistical profile matching the experimental setup it  goes through, but none of them has the statistical profile required by the conjunction of the three predictions. Moreover, none of the sub-collections is statistically similar with any  other sub-collection. Still, the outcomes of experiments are consistent with quantum predictions. 
 The problem is that the sampling is not random. Somehow, the choice of which photon pairs to send to which experimental setup is correlated with the choice of the experimental setup itself. In this case, equations (4)---(6) do not  hold for the entire collection or any particular sub-collection, and $\gamma+\delta+\beta+\gamma$ is larger than or equal to $\beta+\delta$  without contradicting  quantum predictions. In this case, $100\geq 50$; no contradictions exist between outcomes of actual experiments and the assumption of Locality. 

Such a violation of statistical independence would seem to require  some extraordinary conspiracies in Nature.  Not only does this have to be true for these particular setups, which is incredible already, we need there to be similar conspiracies for every such experimental setup, done by anyone, anywhere, and anytime. No matter where, when, and who to carry out the experiment, the strategy requires that no matter what random sampling method we use, the photon pairs with the ``right'' statistical profile should always find themselves at the ``right'' experimental setup. The randomization can be done by a deterministic device that decides, based on the digits of $\pi$, which photon pair goes into which sub-collection. The randomization can also use other mundane methods, such as the rolling of dice, flipping of coin, and the English letters in Act V of  \emph{Hamlet}. No matter what randomization method is used in experiment, the superdeterministic theory will require violations of statistical independence in such a way that the sub-collections will be statistically dissimilar to each other, rendering equations (4)---(6) false of each sub-collection and the entire collection. 
Nature conspires to hide its locality from us. 

Such extraordinary features may be difficult to achieve in any realistic physical theories. Are there any physical theories that can do this? I am not aware of any worked out theory at the moment. However, some initial steps have been taken to investigate possible dynamics and toy models of superdeterminism. \cite{hooft2014cellular} provides an illustration. \cite{hossenfelder2020rethinking} provide  an up-to-date overview and some philosophical discussions.  (\cite{sep-qm-retrocausality} review some ``retrocausal'' models that use backward-in-time causal influences.)

Superdeterminism faces many objections. An important  criticism  focuses on the fact that  endorsing violations of statistical independence would be bad for science in one way or another. After all, the assumption of statistical independence is integral to ordinary statistical inferences.  \cite{shimony1976comment} argue that rejecting statistical independence would undermine the scientific enterprise of discovery by experimentation: 

\begin{quotation}
	In any scientific experiment in which two or more variables are supposed to be randomly selected, one can always conjecture that some factor in the overlap of the backwards light cones has controlled the presumably random choices. But, we maintain, skepticism of this sort will essentially dismiss all results of scientific experimentation. Unless we proceed under the assumption that hidden conspiracies of this sort do not occur, we have abandoned in advance the whole enterprise of discovering the laws of nature by experimentation.
\end{quotation}
Similarly, \cite{MaudlinYoutube} suggests that rejecting it would make it impossible to do science: 

\begin{quotation}
	If we fail to make this sort of statistical independence assumption, empirical science can no longer be done at all. For example, the observed strong robust correlation between mice being exposed to cigarette smoke and developing cancer in controlled experiments means nothing if the mice who are already predisposed to get cancer somehow always end up in the experimental rather than control group. But we would regard that hypothesis as crazy. 
\end{quotation}
These objections based on scientific methodology seem quite compelling to many people.  

Recently, \cite{hossenfelder2020rethinking}  argue that there are multiple mistakes in this type of criticism. One of the mistakes is ``the idea that we can infer from the observation that Statistical Independence is useful to understand the properties of classical systems, that it must also hold for quantum systems. This inference is clearly unjustified; the whole reason we are having this discussion is that classical physics is \emph{not} sufficient to describe the systems we are considering'' (emphasis original).  We may have justification for applying statistical independence to classical systems such as experimental setups involving mice and cigarette smoke. But it does not logically entail that we have justification for applying it to quantum systems of photons and electrons. (What kind of justification do they mean here? I think they mean both epistemic and pragmatic justifications but the text is ambiguous.) 

Their response does not seem to address the worry about scientific methodology. Statistical independence is not the kind of principles we try to empirically justify. Rather, it is part of the inductive principles that we presuppose in order to do science. That is, statistical independence is a precondition for empirical investigation by experimentation. It is not clear what would be an experiment that confirms or disconfirms it, and we may need to assume statistical independence to draw conclusions from the very experiment itself. It may be impossible to empirically justify statistical independence, but that does not suggest there is a problem for applying it in the first place. This follows from a more general observation that even if we cannot empirically justify induction, we are justified in using induction to learn about the world. (See \cite{sep-induction-problem} on Hume's problem of induction.) Hence, their response does not seem to answer the objections of \cite{shimony1976comment} and \cite{MaudlinYoutube}.   

Nevertheless, their response raises an interesting possibility. It is certainly logically consistent for a defender of superdeterminism to maintain that while small microscopic systems (such as electrons and photons) violate statistical independence, large macroscopic systems (such as mice) do not violate it for all practical purposes. That is, we may have reasons to think that the violations of statistical independence may be  suppressed when we reach the macroscopic level. Hence, it is logically consistent for one to claim that statistical independence is false about microscopic systems but for all practical purposes true of macroscopic systems. In short, in ordinary situations when we experiment with mice, we can still use statistical independence; but we should not assume statistical independence  when experimenting with electrons and photons (and other microscopic systems). 

That is of course logically consistent. But we may ask what reasons do we have for thinking that it is true in the superdeterministic theory? One might appeal to decoherence as the mechanism for suppressing certain quantum effects from manifesting in the macroscopic domain (for more on decoherence, see Crull's article in this volume). But decoherence does not fit naturally in a superdeterministic theory.  For one thing, decoherence is primarily about the behaviors of  quantum states (represented by wave functions). However, typically a superdeterministic theory (such as the type favored by \cite{hossenfelder2020rethinking})  does not regard the quantum states to be objective and does not postulate quantum states in the fundamental ontology. Moreover, it is unclear how decoherence can suppress violations of statistical independence. Decoherence  explains the dynamical features that  certain ``branches'' of the wave function do not interfere much with each other. Although the possibility is interesting, there is much work to be done to demonstrate its plausibility in a superdeterministic framework.

%the idea is that the above objections are drawing intuitions for statistical independence from observations about classical systems, such as experimental setups involving mice and cigarette smoke. We do not have conclusive evidence that statistical independence applies to both classical systems as well as quantum systems. Even if we 

%\cite{hossenfelder2020rethinking} respond that while elementary particles have to violate statistical independence for the Bell-type experiments, it does not follow that all  kinds of systems in every context  have to violate statistical independence. 

%As far as I can understand it, the proposal is to limit the violations of statistical independence to particular special systems---those that are subject to Bell-type tests. While this may save the standard method of randomized control trials and experimentation in medical contexts, it seems to be extremely \emph{ad hoc}. But that does not mean it is \emph{absolutely impossible} in Nature. 

 I would like to raise a different worry about superdeterminism. We may worry that superdeterminism of this sort is unlikely to result in a simple fundamental theory. (Here, by ``unlikely'' I mean unlikely in the epistemic sense: unlikely given what we know so far and absent any explicit empirically adequate models that show otherwise.) The constraints on empirical frequencies are so severe that it is hard to see how it can  be written down in any simple formula. (See \cite{kronz1990hidden} for a related argument. See \cite{lewis2006conspiracy} for a discussion of Kronz's argument as well as a new ``measurement problem'' for superdeterminism.)  In order for the local theory to be compatible with the predictions of quantum mechanics, it would have to radically constrain the  state space of the local theory so that only a very small class of histories will be allowed. (Such a constraint can be a joint effect of some lawlike initial conditions and the dynamical laws.)  Not all arrangements of the local parameters will be permitted---otherwise one cannot guarantee perfect agreement  with quantum predictions. What kind of constraints? They will have to encode as much information as the setup and non-local correlations. For example, they would need to entail that an experiment done today using randomization method based on the digits of $\pi$ will somehow still result in statistically dissimilar sub-collections in such a way that produce the desired outcomes of experiments done at arbitrarily far away locations. Similarly it will be the case for randomization based on the letters of Act V of \emph{Hamlet}, the Chinese characters in the \emph{Analects}, or the hexagrams of \emph{I Ching}. No matter what randomization method we choose, the superdeterministic mechanism must ensure that the chosen sub-collection is somehow just the right one for a particular experimental setup. Since the randomization methods seem to have nothing in common,  it is hard to see how the constraints on initial conditions and dynamics can be simple at all. These give us reasons to think that they will be quite complicated.

A defender of superdeterminism may reply that there is a simple formula: just write down the usual Born rule of quantum mechanics and demand that the superdeterministic theory more or less respects that.  It is not clear how to state the Born rule as a simple law in terms of objects accepted on  superdeterminism. As mentioned earlier,  typically a superdeterministic theory (such as the type favored by \cite{hossenfelder2020rethinking}) does not postulate quantum states (represented by wave functions) in the fundamental ontology. After all, a non-separable quantum state may lead to non-local dynamics. However, the Born rule is stated in terms of the quantum state. Respecting the Born-rule statistics (or something close to it) is certainly a nice goal when trying to  construct a local superdeterministic theory with a well-defined ontology and dynamics. The goal is simple (respect the Born rule where it is valid), but it does not follow that the underlying theory will be simple.

Because of the lack of simplicity, the constraints we need to impose in a superdeterministic theory will not look lawlike. Hence, such a theory can be quite complex and difficult to compete with other candidate theories that are far simpler. For example,  \cite{hooft2014cellular}'s Cellular Automaton Interpretation requires the selection of an initial state of the universe, which may be extremely detailed and not at all simple.  Here I take simplicity as a hallmark of fundamental laws of nature. A superdeterministic theory will likely postulate an extremely complicated initial condition (or complicated dynamical laws) that looks nothing like a fundamental law.

Hence, this  problem of superdeterminism  boils down to a violation of a familiar constraint on fundamental laws of nature. A fundamental law should not be too complex. When we evaluate competing theories we are judging them (in part) by the relative complexities of the fundamental laws.  Among competing observationally equivalent theories, the more complex a theory is the lower prior probability we should assign to it. This corresponds to an objective Bayesian way of thinking about  probabilities. However, complexity and simplicity  come in degrees. Now, simplicity and complexity are notoriously vague. But they are indispensable theoretical tools when we confront observationally equivalent theories. For our purpose here, one can plug in any reasonable notion of simplicity and complexity for evaluating scientific theories.  

 In fact, some good physical theories do constrain initial states in order to explain certain wide-spread regularities. For example, in a universe with wide-spread temporal asymmetries, we postulate a low-entropy initial condition. That is now called the Past Hypothesis (\cite{albert2000time}).    We ought to subject the Past Hypothesis  to the constraint of simplicity because it is a candidate fundamental law. It is a candidate fundamental law because it underlies many nomological generalizations such as the Second Law of Thermodynamics and does not seem to be further explained by the dynamics. (This will no longer be true if \cite{carroll2004spontaneous}'s model can successfully explain time's arrow.)
 Fortunately, we have reasons to think that the Past Hypothesis is not  extremely complex. Indeed, it can be specified in terms of simple macroscopic variables (the values of the pressure, density, volume, and energy of the early universe). In certain frameworks, it can even be specified in simple microscopic variables, such as \cite{penrose1979singularities}'s Weyl Curvature Hypothesis or \cite{ashtekar2016initial}'s initial condition for Loop Quantum Cosmology. In the density-matrix-realist framework, the Past Hypothesis can be replaced by the Initial Projection Hypothesis (\cite{chen2018IPH}) that pins down a unique quantum microstate of the universe. It is interesting that  a simple postulate about the initial condition of the universe can explain the wide-spread temporal asymmetries. Part of the reason is due to the structure of state space: there is an asymmetry of macrostate volumes (or dimensions) that emerges as a result of simple dynamics; it is  part of the answer to the problem of time's arrow. Moreover, the Past Hypothesis explanation is perfectly compatible with statistical independence.  

We have good reasons to think that  superdetermistic theories, in contrast, will  postulate something much more complicated than the Past Hypothesis as an initial condition. If such a superdeterministic theory is devised, we should also interpret its initial condition as a fundamental law of nature. (At the very least, it should be given a fundamental axiomatic status in the theory since it is not derived from other laws of the theory.) The wide-spread violations of Bell-type inequalities cry out for explanations. In such a superdeterministic theory, the initial condition is supposed to do the work of explaining why arbitrarily far away events are correlated with each other. We see no reason at all why such a theory (and especially its constraint on the state space) will be simple enough. At least we do not have any evidence that it will be simpler than the competing non-superdeterministic and non-local theories that are already on the market, such as Bohmian mechanics and GRW theory (see the survey articles by Tumulka and Lewis in this volume). 

Hence, there are significant differences between the superdeterministic theory that constrains its initial states to explain Bell-type correlations and a regular quantum theory that constrains its initial states (by the Past Hypothesis) to explain temporal asymmetries. However, these are  differences  in degrees and not of kind.  If a superdeterministic theory aims to recover all quantum predictions, then it would be observationally equivalent to Bohmian mechanics and  more or less equivalent to some versions of GRW theory. But we have good reasons to think that Bohmian mechanics and GRW theory are far simpler than the superdeterministic theory. Hence, the superdeterministic theory should receive much lower prior probability than either Bohmian mechanics or GRW theory. 

%(As for the Past Hypothesis, it does not have any observationally equivalent competitors.  Even if we consider \cite{carroll2004spontaneous}'s model that aspires to explain temporal asymmetries without the Past Hypothesis, it is not clear whether their model is simpler than one with the Past Hypothesis.) 

Nevertheless, that does not mean we should assign 0 credence to superdeterminism. Instead, I think we should follow \cite{bell1977free} and be open-minded in a qualified way: 

%We say predictions 1-3 are true of the collection of photon pairs. But one cannot measure them together. For one thing, some photons will get absorbed. So what we do is to test them by dividing into three sub-collections, A, B, and C. But one way for statistical independence to fail is for there to be some ``preprogramming'' such that the A happens to have statistical characteristics of prediction 1 but not  2 or 3, B with 2 but not  1 or 3, and C with 3 but not 1 or 2. So there fails to be a truly random sampling process. In this way, the collection does not have to satisfy 1 and 2 and 3 together. It just has to satisfy the respective ones in the sub-collections.  Observations will tell us they are fine with quantum predictions, whenever we make the relevant observations --- the right statistic sub-collection will be measured. This theory can be perfectly local, satisfying predetermined values without violating Bell inequalities. In our example in \S2, without stat independence, we cannot deduce (4), (5), and (6), because the collection as a whole does not need to satisfy any of Predictions 1-3, even though it agrees with quantum predictions whenever we make the measurements, and we cannot generate the violation of (3) in (7). 

\begin{quotation}
	Of course it might be that these reasonable ideas about physical randomizers  are 
just wrong -- for the purpose at hand. A theory may appear in which such conspiracies
inevitably occur, and \emph{these conspiracies may then seem more digestible than
the non-localities of other theories.} When that theory is announced I will not refuse
to listen, either on methodological or other grounds. (my emphasis)
\end{quotation}
If one constructs  an empirically adequate superdeterministic theory that is simpler than a non-local theory such as Bohmian mechanics or GRW theory, we should be  be open to assign much higher credence in it. At the moment, no such theory is available. 

\section{Conclusion}

In this short survey article, I introduced Bell's theorem by discussing a simple example. I focused on two strategies that attempt to avoid the conclusion of non-locality: (1) changing the axioms of classical probability theory and (2) embracing superdeterminism and allowing systematic violations of statistical independence. Both have to do in some way with the philosophy of probability. Neither seems promising. Nevertheless, understanding these ideas can help us come to a deeper understanding of Bell's theorem, its significance, and the relevance (or irrelevance) of  the nature of probability.

\section{Note Added}

The article was completed in July 2020. Since then,  an admirably clear article written by G. S. Ciepielewski, E. Okon, and D. Sudarsky (2020) has been posted on arXiv. Here I comment on some of its   features that are relevant to the point I made in \S4. They present a superdeterministic model that exactly reproduces the quantum predictions with a set of strikingly simple dynamical laws and initial condition laws. The theory simulates the whole universe locally, by adding a copy of the universe (an internal space) at each point in physical space, and by stipulating a ``pre-established harmony'' that at $t_0$ the copies of the universe look exactly the same at different points in physical space. How things move  in physical space derive from how things move  in the internal spaces. In the internal spaces, things move according to Bohmian dynamics. Since each internal space occupies only a point in physical space, the fundamental dynamics is  local from the perspective of physical space. 
As the authors acknowledge, their model shares features with Leibniz's Monadology. Hence, I think it is appropriate to call it \textit{Leibnizian quantum mechanics} (LQM). 

(More precisely, the idea of LQM is to (1) take a Bohmian universe of $N$ particles moving in physical space represented by $\mathbb{R}^3$, (2) at each point $x$ in physical space add an internal ``configuration space'' of $\mathbb{R}^{3N}$, (3) replace the universal wave function in Bohmian mechanics with a continuous infinity of wave functions, each defined separately in an internal space and each obeying the Schr\"odinger equation, (4) remove the $N$  particles in physical space from the fundamental ontology, (5) in each internal space, add a point representing the actual configuration in the internal space, whose history depends on the wave function defined in the internal space via a guidance equation, (6) specify a mass density in physical space from the configurations in the internal spaces, and (7) define a simple set of initial condition laws for the ``pre-established harmony'': the initial configurations are the same in all internal spaces and the initial wave functions are the same in all internal spaces, which can be expressed by two simple differential equations.)

The authors themselves acknowledge that their model is not a serious competitor to realist non-local quantum theories such as  standard Bohmian mechanics. Nevertheless, we may wonder what the principled grounds are for its rejection. First,  suppose we understand each internal space of LQM as representing the configuration space of some internal 3-dimensional space at a point in physical space.  Then, ``inside'' each internal space, there are $N$ particles moving according to the usual Bohmian dynamics. If Alice lives in a world described by LQM, there are infinitely many exact copies of her,  of which all except one are made out of particles moving in internal spaces.  Hence, if Alice's self-locating credence is not too biased towards the exceptional one (that she  is made out of particles moving in physical space), she would reason that most likely she lives in an internal space, whose dynamics is non-local. 
Second, it may  be unclear what the fundamental physical space of LQM is. One could stipulate that it is just the physical space. However, if one is sympathetic to \cite{AlbertEQM}'s point that it is something to be inferred and not stipulated, one  runs into a difficulty. In the standard Bohmian case, one can adopt \cite{ChenOurFund}'s criterion and infer that the fundamental physical space is the 3-dimensional physical space. But in LQM, Chen's criterion suggests it is the internal 3-dimensional spaces that should be regarded as physically fundamental, since it is the smallest space that allows a natural definition of the ``multi-field.'' Hence, on some conception, inside the physically fundamental spaces, the dynamics is still non-local. 

Finally and most importantly, even though LQM does not require overly complex initial condition laws or dynamical laws, its ontological additions make it much more complex than standard versions of Bohmian mechanics and GRW theory.   LQM requires the addition of infinitely many more  ``universes'' in addition to the physical space (and to whatever other standard internal spaces we need to postulate). The strictly additional universes, though ``small,'' are exact copies of the physical space including all of its minute details. Unlike the ``emergent'' universes in the Everettian many-worlds interpretation, the infinitely many universes in LQM have a fundamental status. However,  there is no such need to enlarge the ontology on standard Bohmian mechanics or GRW. The only advantage of a superdeterministic theory is its locality, which could be an advantage for seeking a fully relativistic theory.  Unfortunately, in the case of LQM, it is local but it  still requires a preferred rest frame,  which disqualifies it from being fully relativistic. Locality without relativity does not compensate for the increase in complexity. The complexity of LQM lies not in its ``nomology'' (laws) but in its ontology. Hence, LQM is not an empirically adequate theory that is overall simpler and more attractive than a non-local theory such as Bohmian mechanics or GRW theory. The final point of \S4 still stands. 

Nevertheless, LQM is a rare example of an explicit  superdeterministic model of the universe that reproduces the exact predictions of quantum mechanics. It adds a great deal of clarity for understanding the relative costs and benefits of maintaining locality and rejecting  statistical independence. One may try to avoid complicating the laws by complicating the ontology instead; but either way one has to complicate some part of the theory. That could well be a generic feature of any superdeterministic theory that attempts to avoid the charge of non-locality.

\section{Further readings}

For discussions of the issue of ``realism'' in Bell's proof, see  T. Norsen, ``Against `realism''', \textit{Foundations of Physics}, 37(3):311-340, 2007, T. Maudlin, ``What Bell did'', \textit{Journal of Physics A: Mathematical and Theoretical}, 47(42):424010, 2014, and R. Tumulka, ``The assumptions of Bell's proof'', in M. Bell and S. Gao (eds.), \textit{Quantum Nonlocality and Reality: 50 Years of Bell's Theorem} (Cambridge University Press, 2016). 
For discussions of non-locality, superluminal signaling, and relativistic invariance, T. Maudlin's \textit{Quantum non-locality and relativity: Metaphysical intimations of modern physics} (Wiley, 2011) is a landmark monograph on the topic; for collapse models that demonstrate the compatibility of Lorentz invariance and non-locality, see  R. Tumulka, ``A relativistic version of the Ghirardi-Rimini-Weber model'', \textit{Journal of Statistical Physics}, 125(4): 821-840, 2006, and D. Bedingham et al., ``Matter density and relativistic models of wave function collapse'', \textit{Journal of Statistical Physics}, 154(1-2): 623-631, 2014. 
For discussions of locality and non-locality in the many-worlds interpretation of quantum mechanics, see D. Wallace, \textit{The Emergent Multiverse: Quantum theory according to the Everett interpretation} (Oxford University Press, 2012), and V. Allori et al., ``Many worlds and Schr\"odinger's first quantum theory'', \textit{British Journal for the Philosophy of Science}, 62(1):1-27, 2010. 
For discussions of parameter independence and outcome independence, see J. P. Jarrett, 
	``On the physical significance of the locality conditions in the Bell arguments'', \textit{No\^us}, 18(4) 569-589, 1984, A. Shimony, ``Search for a worldview which can accommodate our knowledge of microphysics'', in J. T. Cushing and E. McMullin (eds.), \textit{Philosophical Consequences of Quantum Theory} (University of Notre Dame Press, 1989), 62-76, R. Healey, ``Chasing quantum causes, how wild is the goose?'' \textit{Philosophical Topics}, 20(1):181-204, 1992, and T. Maudlin's \textit{Quantum non-locality and relativity: Metaphysical intimations of modern physics}, Ch.4. 
	For discussions of causation and causal explanations, see J. S. Bell, ``Bertlmann's socks and the nature of reality,'' \textit{Le Journal de Physique Colloques}, 42(C2): C2-41, 1981, M. L. Redhead, ``Nonfactorizability, stochastic causality, and passion-at-a-distance'', in J. T. Cushing and E. McMullin (eds.), \textit{Philosophical Consequences of Quantum Theory} (University of Notre Dame Press, 1989), 145-153,  R. Healey, ``Chasing quantum causes, how wild is the goose?'', and T. Maudlin's \textit{Quantum non-locality and relativity: Metaphysical intimations of modern physics}, Ch.5. 
	For a survey of experimental tests and certain loophole-free tests of Bell's inequalities, see \S4-\S5 of W. Myrvold and A. Shimony, ``Bell's theorem'', \textit{Stanford Encyclopedia of Philosophy} (2019). 
	For a discussion of related matters from a relativity-centered perspective, see Chapter 2d by W. Myrvold in this volume. 
%	\item Retrocausality: see \cite{sep-qm-retrocausality} for a review. 

\nocite{norsen2007against, maudlin2014bell, tumulka2015assumptions, maudlin2011quantum, tumulka2006relativistic, bedingham2014matter, wallace2012emergent, allori2010many, jarrett1984physical, shimony1989search, healey1992chasing, bell1981bertlmann, redhead1989nonfactorizability, healey1992chasing, maudlin2011quantum,  sep-bell-theorem, ciepielewski2020superdeterministic}

%------------------------------------------------

\section*{Acknowledgement}
I am grateful for helpful discussions with Craig Callender, Sheldon Goldstein, Mario Hubert, and Isaac Wilhelm, as well as written comments from  Alan H\'ajek, Sabine Hossenfelder, Kelvin McQueen, Peter Morgan, Travis Norsen, Elias Okon, Timothy Palmer, and Alastair Wilson.

%David Albert, Craig Callender, Sheldon Goldstein, Jill North, Ted Sider, 

%----------------------------------------------------------------------------------------
%	BIBLIOGRAPHY
%----------------------------------------------------------------------------------------

\bibliography{test}

%----------------------------------------------------------------------------------------

\end{document}